%
%
%
%
%
%
%
%
%
%
%
%
%
%
\input phyzzx
\input epsf
%
%
\catcode`\@=11 
\def\papersize{\hsize=40pc \vsize=53pc \hoffset=0pc \voffset=1pc
   \advance\hoffset by\HOFFSET \advance\voffset by\VOFFSET
   \pagebottomfiller=0pc
   \skip\footins=\bigskipamount \normalspace }
\catcode`\@=12 
\papers
\newcount\figno
\figno=0
\def\fig#1#2#3{
\par\begingroup\parindent=0pt\leftskip=1cm\rightskip=1cm\parindent=0pt
\baselineskip=11pt
\global\advance\figno by 1
\midinsert
\epsfxsize=#3
\centerline{\epsfbox{#2}}
\vskip 12pt
{\bf Fig. \the\figno:} #1\par
\endinsert\endgroup\par
}
\def\figlabel#1{\xdef#1{\the\figno}}
\def\encadremath#1{\vbox{\hrule\hbox{\vrule\kern8pt\vbox{\kern8pt
\hbox{$\displaystyle #1$}\kern8pt}
\kern8pt\vrule}\hrule}}

\def\pp{$(n_e,n_m)$}
\def\to{\rightarrow}

\def\ad{a_D}
\def\e{\epsilon}


\vsize=23.cm
\hsize=15.cm
\def\IM{\mathop{\Im m}\nolimits}

\def\Z{{\bf Z}}
\def\ZZ{$\Z_2$}
\def\R{{\bf R}}
\def\H{{\bf H}}
\def\CC{${\bf C}^2$}
\def\C{${\cal C}$}
\def\CP{${\cal C}^+$}
\def\CM{${\cal C}^-$}

\tolerance=500000
\overfullrule=0pt

\Pubnum={LPTENS-96/16 \cr
{\tt hep-th@xxx/9602082} \cr
February 1996}

\date={}
\pubtype={}
\titlepage
\title{\bf  The Strong-Coupling Spectrum of the Seiberg-Witten Theory}  
\author{Frank~Ferrari}
\andauthor{Adel~Bilal}
\vskip 1.cm
\address{
CNRS - Laboratoire de Physique Th\'eorique de l'\'Ecole
Normale Sup\'erieure
\foot{{\rm unit\'e propre du CNRS, associ\'ee \`a l'\'Ecole Normale
Sup\'erieure et l'Universit\'e Paris-Sud.}}
 \nextline 24 rue Lhomond, 75231
Paris Cedex 05, France\break
{\tt ferrari@physique.ens.fr,\  bilal@physique.ens.fr}
} 

\vskip 1.cm
\abstract{We carefully study the global structure of the solution
of the $N=2$ supersymmetric pure Yang-Mills theory with gauge group 
$SU(2)$ obtained by Seiberg and Witten. We exploit its \ZZ-symmetry and
describe the curve in moduli space where BPS states can become unstable, 
separating the strong-coupling from the weak-coupling region. This 
allows us to obtain the spectrum of stable BPS states in the 
strong-coupling region: we prove that only the two particles responsible for the 
singularities of the solution (the magnetic monopole and the dyon of unit
electric charge) are present in this region. Our method also permits us to 
very easily obtain the weak-coupling spectrum, without using 
semi-classical methods. We discuss how the BPS states disintegrate 
when crossing the border from the weak to the strong-coupling region.}

\endpage
\pagenumber=1

 \def\PL #1 #2 #3 {Phys.~Lett.~{\bf #1} (#2) #3}
 \def\NP #1 #2 #3 {Nucl.~Phys.~{\bf #1} (#2) #3}
 \def\PR #1 #2 #3 {Phys.~Rev.~{\bf #1} (#2) #3}
 \def\PRL #1 #2 #3 {Phys.~Rev.~Lett.~{\bf #1} (#2) #3}
 \def\CMP #1 #2 #3 {Comm.~Math.~Phys.~{\bf #1} (#2) #3}
 \def\IJMP #1 #2 #3 {Int.~J.~Mod.~Phys.~{\bf #1} (#2) #3}
 \def\JETP #1 #2 #3 {Sov.~Phys.~JETP.~{\bf #1} (#2) #3}
 \def\PRS #1 #2 #3 {Proc.~Roy.~Soc.~{\bf #1} (#2) #3}
 \def\JFA #1 #2 #3 {J.~Funkt.~Anal.~{\bf #1} (#2) #3}
 \def\LMP #1 #2 #3 {Lett.~Math.~Phys.~{\bf #1} (#2) #3}
 \def\IJMP #1 #2 #3 {Int.~J.~Mod.~Phys.~{\bf #1} (#2) #3}
 \def\FAA #1 #2 #3 {Funct.~Anal.~Appl.~{\bf #1} (#2) #3}
 \def\AP #1 #2 #3 {Ann.~Phys.~{\bf #1} (#2) #3}
 \def\MPL #1 #2 #3 {Mod.~Phys.~Lett.~{\bf #1} (#2) #3}

\REF\SW{N. Seiberg and E. Witten, {\it Electric-magnetic duality, monopole
condensation, and confinement in $N=2$ supersymmetric Yang-Mills theory}, 
\NP B426 1994 19 , {\tt hep-th/9407087}.}

\REF\MAT{P.C. Argyres and A.D. Shapere, {\it The vacuum structure of $N=2$ 
super QCD with
classical gauge groups}, Rutgers preprint RU-95-61, {\tt hep-th/9509175}.}

\REF\OS{H. Osborn, {\it Topological charges for $N=4$ supersymmetric
gauge theories and monopoles of spin 1},
\PL B83 1979 321 .}

\REF\SWW{N. Seiberg and E. Witten, {\it Monopoles, duality and chiral
symmetry breaking in $N=2$ supersymmetric QCD}, \NP B431 1994 484 ,
{\tt hep-th/9408099} .}

\REF\MO{C. Montonen and D. Olive, {\it Magnetic monopoles as gauge particles?}, 
\PL B72 1977 117 .}

\REF\GAU{J.P. Gauntlett, {\it Low-energy dynamics of $N=2$
supersymmetric monopoles}, \NP B411 1994 443 \hfill\break
A. Sen, {\it Dyon-monopole bound states, self-dual harmonic forms
on the multi-monopole moduli space, and $SL(2, \Z)$ invariance
in string theory}, \PL B329 1994 217 .}

\REF\VAF{S. Cecotti, P. Fendley, K. Intriligator and C. Vafa,
{\it A new supersymmetric index}, \NP B386 1992 405 \hfill\break
S. Cecotti and C. Vafa, {\it On classification of $N=2$
supersymmetric theories}, \CMP 158 1993 569 .}

\REF\FAY{A. Fayyazuddin, {\it Some comments on $N=2$ supersymmetric
Yang-Mills}, \MPL A10 1995 2703 , {\tt hep-th/9504120}.}

\REF\BIL{A. Bilal, {\it Duality in $N=2$ susy $SU(2)$ Yang-Mills theory: 
A pedagogical introduction to
the work of Seiberg and Witten}, \'Ecole Normale Sup\'erieure preprint LPTENS-95/53,
{\tt hep-th/9601007}.}

\REF\SEI{N. Seiberg, {\it Supersymmetry and non-perturbative beta functions},
\PL B206 1988 75 .}

\REF\ERD{A. Erdelyi et al, {\it Higher Transcendental Functions}, Vol 1, McGraw-Hill,
New York, 1953.}

\REF\MATON{M. Matone, {\it Koebe $1/4$-theorem and inequalities in
$N=2$ super-QCD}, Padua preprint DFPD-95-TH-38, {\tt hep-th/9506181}.}

\REF\ARG{P.C. Argyres, A.E. Faraggi and A.D. Shapere, {\it Curves
of marginal stability in $N=2$ super-QCD}, preprint IASSNS-HEP-94/103, 
UK-HEP/95-07, {\tt hep-th/9505190}.}

\REF\ROC{U. Lindstr\"om, M. Ro\v cek, {\it A note on the
Seiberg-Witten solution of $N=2$ super Yang-Mills theory},
\PL B355 1995 492 , {\tt hep-th/9503012}.}

\REF\BF{A. Bilal and F. Ferrari, {\it Curves of marginal stability, and weak and
strong-coupling BPS spectra in $N=2$ supersymmetric QCD}, preprint LPTENS-96/22,
HUB-EP-96/11, {\tt hep-th/9605101}.}

\chapter{Introduction}
In an already classical paper, Seiberg and Witten [\SW] derived
the exact low energy wilsonian effective action for the pure $N=2$ supersymmetric
Yang-Mills theory with gauge group $SU(2)$. (Note that pure gauge $SU(2)$, i.e. without
extra matter, is actually equivalent to $SO(3)$.) Since then,
their work has been generalized to other gauge groups and to
theories with matter (see, {\it e.g.} [\MAT] and references
therein). The main ingredient that allows for an exact solution
of these strongly interacting theories is duality: one
can  perform certain duality transformations (the duality
group being $Sp(2,{\Z})\equiv SL(2,{\Z})$ in the case of an $SU(2)$ theory)
which relate different descriptions of the same low energy theory in terms of 
different sets
of elementary fields. The latter are mutually
non local. In this sense, a duality
transformation is like a change of variables, not a symmetry of the
theory. However, in some particular models, like
$N=4$ super Yang-Mills theory [\OS] or $N=2$ super  $SU(2)$ Yang-Mills theory
with four flavors [\SWW], these duality transformations are conjectured
to be
a symmetry of the theory (Montonen-Olive duality [\MO]). 
In particular, the spectrum of BPS saturated
states should then be self-dual. Moreover, even in those theories in which
Montonen-Olive duality does not hold in full, there may exist some
class of duality transformations under which the spectrum is
invariant. For instance, it was argued in [\SW] that the semiclassical
({\it i.e.} the weak-coupling) spectrum should be invariant under the
transformation which
corresponds to the monodromy around the point
at infinity in the moduli space, \penalty -1000
$(n_e, n_m) \rightarrow (-n_e+2n_m, -n_m)$,\penalty -1000\ where
$n_e$ and $n_m$ are the electric and 
magnetic charges of the stable BPS states.

In the present paper, restricting our attention to the $SU(2)$ case of [\SW], 
we exploit the existence of another duality
({\it i.e.} $Sp(2,{\Z})$) transformation,
related to the \ZZ-symmetry of the
moduli space, under which the spectrum should be invariant.
This, together with a few other arguments insuring the physical consistency
of the exact Seiberg-Witten solution, will allow us to
unambiguously  determine both the weak-coupling spectrum
(which has already been investigated when $n_m \leq 2$ using semi-classical methods
[\GAU]),
and, more important, also the strong-coupling spectrum.
In the weak-coupling region the spectrum of BPS states contains
all dyons $\pm(n,1)$ with unit magnetic charge and arbitrary integer electric charge
$n$, in addition to the perturbative W-bosons $\pm(1,0)$. When entering the
strong-coupling region, almost all of these states decay and one is left with {\it
only two} BPS states, namely the magnetic monopole $\pm(0,1)$ and the 
dyon of unit electric charge
described either as
$\pm(1,1)$ or as $\pm(1,-1)$. This dramatic change in the spectrum 
is possible precisely since the weak and strong-coupling regions in moduli space are
separated by a curve \C\ where the otherwise stable BPS states can become unstable.
Such a phenomenon was first considered in two dimensional theories in
[\VAF]. We
use the above-mentioned \ZZ-symmetry to prove that, for the Seiberg-Witten solution, 
no other BPS states can be present
in the strong-coupling region.

This paper is organized as follows:
Section 2 is a brief review of the work of Seiberg and Witten [\SW] to fix our notation
and insist on the exact \ZZ-symmetry of the theory.
Particular attention is paid to the explicit form of the solution of [\SW ]
and to its analytic structure. In Section 3, we describe the curve
${\cal C}$ on which the usually stable BPS states become unstable and which 
is the border between the strong
and weak-coupling regions of the moduli space ${\cal M}$.
Section 4 is devoted to the \ZZ-symmetry and its consequences.
In Section 5, we then rigorously determine both the weak 
and strong-coupling spectra. 
In Section 6, we present a simple 
physical consistency check of our solution, illustrating how the BPS states 
decay\foot{
Such kinematics of possible decay reactions were also considered in [\FAY], indicating
already the possibility of a strong-coupling spectrum
consisting only of the monopole and the dyon.}
when crossing the curve \C.
Finally, in Section 7, we recapitulate our assumptions and conclusions.

\chapter{Overview of Seiberg-Witten theory}
For most of the material presented in this section, see [\SW] and
references therein (see also [\BIL] for a pedagogical introduction).
The microscopic action $S_{\rm mic}$ of $N=2$ supersymmetric 
gauge theory without
hypermultiplets ({\it i.e.} without extra matter) is expressed in terms of an  
$N=2$ vector superfield  
$\Psi^a$ transforming in the adjoint representation of the gauge group, which
for us will be $SU(2)$, or equivalently $SO(3)$.
Among others, $\Psi$ contains  a scalar field $\phi$ whose
potential is 
$V(\phi)={1\over 2}$tr$(\lbrack\phi ^{\dagger},\phi\rbrack )^2$.
Thus, as long as $\phi$ and $\phi^{\dagger}$ commute in $su(2)$, 
the scalar potential remains
zero even for a  nonvanishing expectation value of $\phi$ which spontaneously breaks the
$SU(2)$ gauge symmetry down to $U(1)$.
This shows that, at least semiclassically, the theory 
has a
continuum of gauge inequivalent vacua, called the moduli space, 
parametrized by the gauge
invariant quantity $u=\langle$tr$\, \phi ^2\rangle$. Seiberg and
Witten argued [\SW] that this picture is maintained quantum
mechanically, $u$ being a good local coordinate on the quantum
moduli space ${\cal M}$.

Among other symmetries, the action $S_{\rm mic}$ has a global $U(1)_R$ 
$R$-symmetry under which 
$\phi$ has charge 2.
This $U(1)_R$ symmetry is reduced by an anomaly down to ${\Z}_8$.
This can  be seen from the form of the instanton contributions in
the low energy effective action
[\SW,\SEI].
Since
$u=\langle$tr$\, \phi ^2\rangle$ has charge 4 under this symmetry, a given vacuum with a 
non-vanishing value
of $u$ furthermore breaks ${\Z}_8$ to ${\Z}_4$.  
Nervertheless, let
us stress that 
$u$ and
$-u$ correspond to physically equivalent vacua related by 
the ${\Z}_8$-symmetry of the quantum
theory. This is the \ZZ-symmetry on the moduli space
which we will extensively use in the following.

As already mentioned, at a generic point $u\in{\cal M}$ the
gauge symmetry is broken down to $U(1)$ by the vacuum expectation value of $\phi$,
$\langle\phi\rangle ={1\over 2}a(u)\sigma _3$. The low energy wilsonian effective
lagrangian $\cal L$  then is expressed 
in terms of 
the light fields of the
microscopic theory. By $N=2$ supersymmetry, the most general form for $\cal L$ is,
in terms of the $N=1$ abelian vector ($W$) and chiral ($A$) superfields,
$${\cal L}={1\over 8\pi}\,\IM\,\left[2\int d^2\theta d^2\overline{\theta}
\,A_D
\overline{A}+\int d^2\theta\, {\cal F}''(A)W^2\right] \ .
\eqn\dii$$
where $\cal F$ is a holomorphic function and
$A_D\equiv{\cal F}'(A)$ is the dual superfield of $A$.
We also note $a_D = {\cal F}'(a)$.
An $Sp(2,{\Z})$
transformation on $\Omega =\bigl(a_D (u),a(u)\bigr)$ is simply a duality transformation under
which $\cal L$ is invariant. 
Then $\Omega $ is naturally interpreted as a section
of  a holomorphic $Sp(2,{\Z})$ vector bundle $E$ over the moduli 
space  $\cal M$, with fiber \CC .
One can then define a symplectic
product $\eta$ of two sections $\alpha =(\alpha _1,\alpha _2)$ and
$\beta =(\beta _1,\beta _2)$
by $\eta (\alpha ,\beta)=\alpha _2\beta _1
-\alpha _1\beta _2$.
In this notation, the K\"ahler potential on $\cal M$ is
$K={i\over 2}\eta (\Omega,\overline\Omega)$ from which the 
K\"ahler metric  is derived as $ds^2=\IM\,
(da_D\,d\overline{a})$.
This is a positive definite metric as a consequence of unitarity.

In this language, a BPS state will be represented by a locally
constant section $p=(n_e,n_m)$ over $\cal M$. The mass of such
a state is :
$$m=\sqrt{2}\,\vert Z \vert\ , \quad Z=\eta(\Omega ,p) =a n_e - a_D n_m \ ,
\eqn\diii$$
where $Z$ is the central charge of the supersymmetry
algebra. The mass $m$, being given by the symplectic product of $\Omega$ and $p$, is
obviously an $Sp(2,\Z)$-invariant.
This remarkable formula shows that once we know the
section $\Omega$ and the set of sections $p$ representing the BPS
particles, we 
also have the mass spectrum of the theory.
Seiberg and Witten completely determined $\Omega$. This will be
the starting point of our analysis.

Let us carefully examine
the explicit form of $\Omega$.
$a_D$ and $a$ can be expressed in terms of hypergeometric functions
as [\BIL]:
$$\eqalign{
\ad(u)&=i{u-1\over 2} F\left({1\over 2},{1\over 2},2;{1-u\over 2}\right)\cr
a(u)&=\sqrt{2} (u+1)^{1\over 2} F\left(-{1\over 2},{1\over 2},1;{2\over u+1}\right)
\ .\cr }
\eqn\div$$
Here the square-root is defined with the argument of a complex
number always running from $-\pi$ to $\pi$. Recall that $F(a,b,c;z)$ has a cut on the
positive real
axis from $z=1$ to $z=+\infty$. Hence $a_D(u)$ has a cut on the real line from 
$-\infty$ to $-1$, while $a(u)$ has two cuts, both on the real line, one from 
$-\infty$ to $-1$ and another from $-1$ to $1$, see Fig. 1.
\vskip 3.mm
\fig{The branch cuts of $a_D(u)$ (left) and of $a(u)$ (right).}{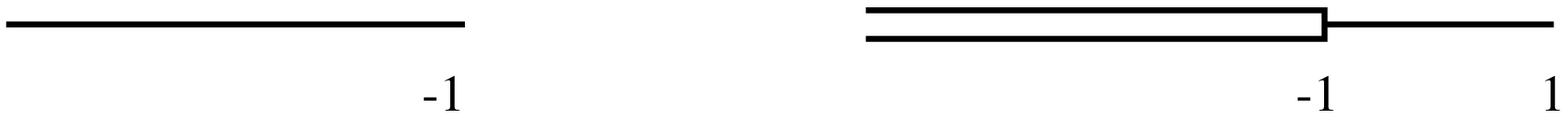}{12cm}
\figlabel\figi
\vskip 2.mm

Near the singular points which are
the branch points $1$ and $-1$ and the point at infinity, 
the asymptotic behaviour of $\Omega$ is:
$$\left.\eqalign{
\ad(u)&\simeq {i\over \pi} \sqrt{2u}\left[ \log u +3 \log 2-2\right] \cr
a(u)&\simeq \sqrt{2u} } \quad \right\} \quad {\rm as\ } u\to\infty$$
$$\left.\eqalign{
\ad(u)&\simeq i{u-1\over 2}\cr
a(u)&\simeq {4\over \pi}-{1\over \pi}{u-1\over 2} \log{u-1\over 2}+
{1\over \pi}{u-1\over 2}\left( -1+4\log 2\right) } \quad 
\right\} \quad {\rm as\ } u\to 1$$
$$\left.\eqalign{
\ad(u)&\simeq  {i\over \pi}\left[  -{u+1\over 2} \log{u+1\over 2} + {u+1\over 2}
\left( 1+4\log 2\right) -4\right]\cr
a(u)&\simeq  {i\over \pi}\left[ \e {u+1\over 2} \log{u+1\over 2} + {u+1\over 2}
\left( -i\pi -\e\left(1+4\log 2\right)\right) +4\e\right]
 } \quad  \right\} \quad {\rm as\ } u\to -1 \ ,
\eqn\dv$$
where $\epsilon$ is the sign of $\IM\, u$. From these formula one can
recover the monodromies associated with the analytic continuations of $\Omega$ around 
the three singular points. Around $\infty$
and $1$ they are:
$$M_{\infty}=\pmatrix{-1 &\hfill2\cr\hfill 0 &-1\cr},\,\, 
M_1=\pmatrix{\hfill 1 &0\cr -2 &1\cr}.
\eqn\dvi$$
Around $-1$, the monodromy depends explicitely on the base point $u_P$
chosen to define the monodromy group. This is due to the appearence of
$\epsilon$ in eq.~\dv . One obtains the matrix $M_{-1}$ if $\IM\, u_P < 0$ and
$M_{-1}'$ if $\IM\, u_P > 0$:
$$M_{-1}=\pmatrix{-1 &2\cr -2 &3\cr},\,\, 
M_{-1}'=\pmatrix{\hfill 3 &\hfill 2\cr -2 &-1\cr}.
\eqn\dvii$$
We have $M_1 M_{-1} = M_{\infty} =M_{-1}' M_{1}$. Note that 
going round a singular point one is bound to cross a cut 
and thus change from the principal branch to another branch
of the multivalued function $(a_D(u),a(u))$. The asymptotics \dv\ and monodromy
matrices \dvi\ and \dvii\ are valid
for the principal branch.  If then one goes around a
second singularity, one obtains a monodromy matrix corresponding to the new branch and
which may 
differ from the ones quoted above by conjugation by the first monodromy matrix. 
This invalidates
simple contour composition
arguments suggested in [\SW] for determining $M_{-1}$ for instance.
Though such subtleties were completely irrelevant in the analysis
of [\SW], they are important here.

We close this section by pointing out that, according to the previous remarks, the
explicit solution \div\ is not the only one compatible with
the physical constraints used to derive it. One may use the analytic
continuations of $a_D$ and $a$  obtained by going
around infinity $p$ times. This amounts to conjugating the representation
of the monodromy group by replacing the monodromy matrices $M$ by
$M_{\infty}^p M M_{\infty}^{-p}$. 
This does not change the asymptotics of $a_D$ and $a$ as $u \to \infty$ which was the
physical input.
We will call this the
democracy transformation because it is at the origin of the 
``democracy'' between dyons as noted in [\SW].

\endpage
\chapter{The curve $\IM\,(a_D/a)=0$} 

As already noted, a BPS state is represented by a locally
constant section $p=(n_e,n_m)$ where $n_e$ and $n_m$ are relatively
prime integers. 
Recall that its mass is 
proportional to the euclidean length of the complex vector
$a(u) n_e-a_D(u) n_m$ (eq.~\diii ). Recall also 
that if $a_D/a$ is not real, the set of these
vectors forms a lattice in the complex plane. If $n_e$ and $n_m$ are not relatively
prime, i.e. if $(n_e,n_m)
= q(n,m)$ for $n,m,q\in {\Z},\ q\ne \pm 1$, then the BPS state is unstable against
decay into $q$ BPS states $(n,m)$ since this reaction conserves the total electric and
magnetic charges as well as the total mass. On the contrary, if $n_e$ and $n_m$ are
relatively prime it follows from the conservation laws that this state cannot decay and
hence is stable as long as $a_D/a\notin {\R}$. On the other hand, if $a_D/a$ is
real, it becomes much easier to satisfy charge and mass conservation, and otherwise
stable BPS states can decay (see {\it e.g}. [\SW]). We will study examples
of such decays in Section 6. 

We define the curve \C\ on the moduli space ${\cal M}$ as
${\cal C}=\bigl\{ u\in {\cal M}\,|\,\IM\, (a_D/a) =0\bigr\}$.
Note that the solution \div\ is such that $a(u)$ never vanishes. In fact, 
for all $u\in {\cal M}$ one has 
$\vert a(u) \vert \ge \vert a(0) \vert \simeq 0.76$. 
This means that the singular point $a=0$ of the classical
moduli space where the full $SU(2)$ gauge symmetry is restored does no longer exist in
the quantum moduli space ${\cal M}$.
The curve \C\ is of utmost importance if one wants to study the spectrum of the theory.
As long as two points $u$ and $u'$ in ${\cal M}$ are not separated by the curve \C,
{\it i.e.} if they can be joined by a continuous path in ${\cal M}$ that does not cross
\C, one can deform the theory at $u$ into the theory at $u'$ without
changing the spectrum. 
By spectrum, we mean the set of locally constant sections representing
BPS particles. Of course the mass spectrum will change as $u$ varies.

To try to determine the curve \C\ analytically one observes [\BIL] that $a_D$ and $a$
are both solutions of the same second-order differential equation
$$
\left[-{{\rm d}^2\over {\rm d}u^2} + V(u)\right] \pmatrix{a_D(u)\cr a(u)\cr} = 0 \ ,
\quad V(u)=-{1\over 4(u^2-1)} \ .
\eqn\tti$$
%
%
%
It is then well-known [\ERD] that the ratio of two solutions of \tti\ satisfies
the Schwarz equation
$$w(u)={a_D(u)\over a(u)} \quad \Rightarrow \quad \{w,u\}= -2\, V(u) \ ,
\eqn\ttiii$$
where $\{w,u\}$ denotes the Schwarzian derivative of $w$ with respect to $u$:
$\{w,u\}={w'''\over w'}-{3\over 2}\left({w''\over w'}\right)^2$. The curve \C\
precisely is the set of points where $w(u)$ is real. Hence a parametrisation of \C\
would be given by the inverse function $u(w)$ with $w$ a parameter in an appropriate
real interval.
Using this line of reasoning, a parametric form of \C\ was obtained
in [\MATON]. The shape of \C\ was also discussed in [\FAY,\ARG] (see
Fig. 2). Actually the precise determination of \C\ is completely
irrelevant for our purposes. The general features which we need are 
easy to check numerically and are summarized below.

\vskip 2.mm
\fig{The curve \C\ in the complex $u$-plane
where $\IM\, (a_D/a)$ vanishes.}{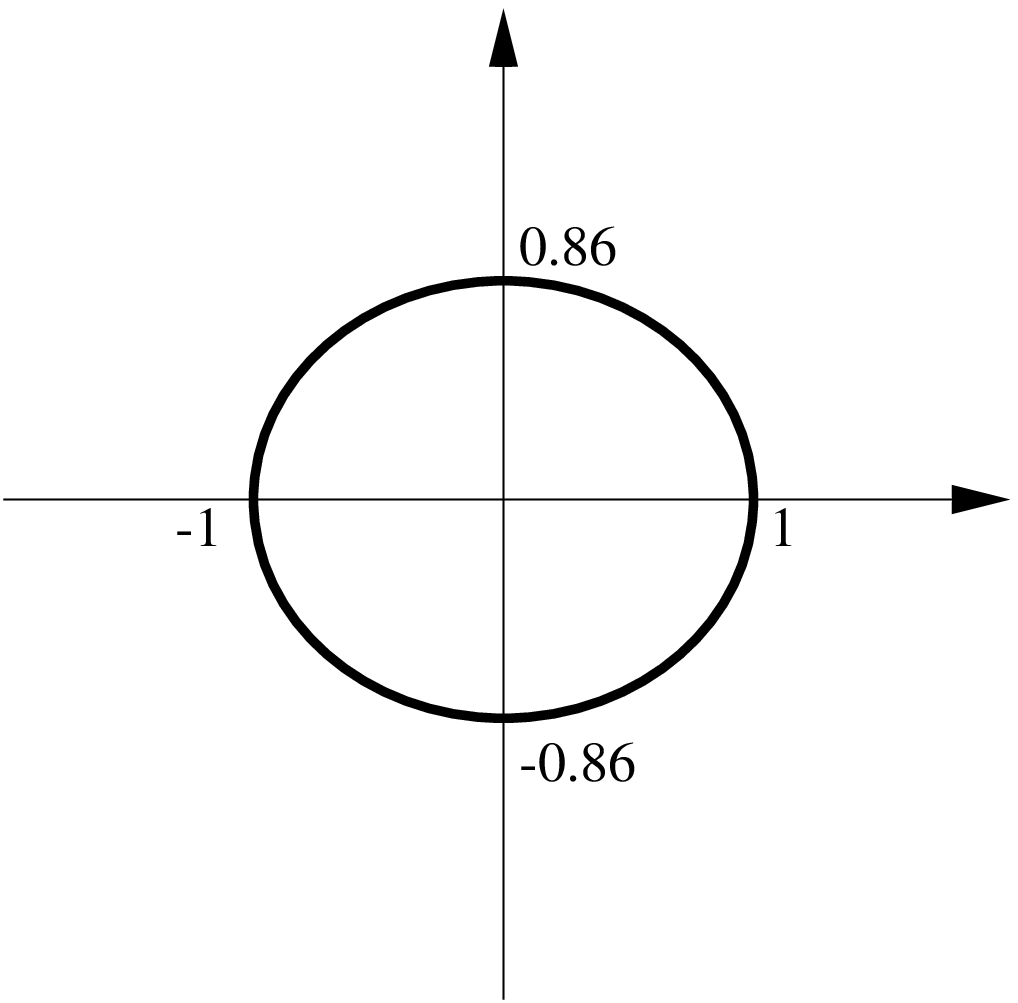}{8cm}
\figlabel\figii
\vskip 2.mm

To a very good approximation (about $10^{-2}$) \C\ looks like an ellipse
(although it is not {\it exactly} an ellipse), centered at the origin of the complex plane and
with semimajor and semiminor axis equal 
respectively to 1 and $\simeq 0.86$. In particular the points $u=\pm 1$ are on the
curve \C. \C\ does not contain any other disconnected
component elsewhere in the complex $u$-plane. 
The curve \C\ being closed, it separates 
a strong-coupling region ${\cal R}_S$ containing $u=0$
from a weak-coupling region
${\cal R}_W$ containing $u=\infty$. The physical spectrum of BPS states may 
change from one region to the other, but it is necessarily the same at any
two points of $\cal M$ in the same region. We will call
${\cal S}_S$ and ${\cal S}_W$ the two a priori different spectra.

Another interesting property of the curve $\cal C$ is that massless BPS states can only
exist on this curve. Indeed, since $m=\sqrt{2}\, \vert a n_e-a_D
n_m\vert$, for a massless state one must have ${a_D\over a}={n_e\over n_m}$ which is
rational, and {\it a fortiori} real. In this respect it is interesting to know which
(real) values ${a_D\over a}$ takes on the curve \C. Let us call \CP\ the part of \C\
in the upper half $u$-plane and \CM\ the part in the lower half-plane.
(We include the two common end-points $u=\pm 1$ into both, \CP\ and \CM.) 
It is easy to see
from the explicit expressions \div\ that on  \CP\ one has ${a_D\over a}
\leq 0$ and 
${a_D\over a}(u)\to -1$ as $u\to -1$, while on \CM\ one has ${a_D\over a}
\geq 0$ and 
${a_D\over a}(u)\to +1$ as $u\to -1$. Furthermore, it is clear that ${a_D\over
a}(1)=0$. This just expresses the well-known result [\SW] that at $u=1$ the magnetic
monopole is massless. At $u=-1$, the massless state is a dyon, but it is described as
$(n_e,n_m)=\pm (1,-1)$ if one approaches $u=-1$ from the upper half-plane, and as 
$(n_e,n_m)=\pm (1,1)$ if the same point is approached from the lower half-plane.
The fact that the same BPS state has two different descriptions is important for us
and will be discussed in more detail below. As one  goes along \CP\ from $u=-1$ to
$u=1$,  ${a_D\over a}$ increases continuously and monotonically from $-1$ to $0$.
Going back from $u=1$ to $u=-1$ on \CM, ${a_D\over a}$ continues to increase
continuously and monotonically from $0$ to $+1$. Again, one can numerically determine
${a_D\over a}(u)$ on the curve \C. 
However, the precise form will
not be important for our analysis below. What is important is that ${a_D\over a}(u)$
can take {\it any} value in the real interval $[-1,1]$ for $u\in$ \C, with
{\it any} value in $[-1,0]$ for $u\in$ \CP\ and 
{\it any} value in $[0,1]$ for $u\in$ \CM.

It follows from the above remarks that a BPS state $(n_e,n_m)$ 
of the weak-coupling spectrum ${\cal S}_W$ will
become massless somewhere on the curve \C\ if ${n_e\over n_m}\in\lbrack -1,
1\rbrack$. 
Now, when a charged particle becomes
massless, their should be a singularity in the effective gauge coupling
and thus 
in the low energy wilsonian
action.
We know that there are precisely two such  singularities at $u=1$ and
$u=-1$, where $a_D/a=\pm 1$ or $0$. From this we conclude that the monopole 
$\pm (0,1)$  as well as the
dyon described either as $\pm (1,1)$ or 
$\pm (1,-1)$ do exist in both ${\cal R}_W$ and ${\cal R}_S$ since these are the only stable
states able to yield these singularities.
We also conclude from the absence of other singularities that there are no other
states in ${\cal R}_W$ or ${\cal R}_S$ that become massless on \C. 
Finally, note that any state $p$ which becomes massless somewhere
on $\cal C$ must exist both in ${\cal R}_W$ and ${\cal R}_S$. Indeed,
first, the singularity produced by such a state can be seen from
the two sides of the curve. Second,
one can cross $\cal C$ precisely at the point where $p$ becomes massless
and it is then stable under decay since it is the only massless charged particle
at that point. This is the case for the monopole and the dyon. 
\chapter{The \ZZ-symmetry}

\section{Global symmetries on the moduli space}

We are now going to examine the realization of the \ZZ-symmetry $u\to u'=-u$ on the
moduli space. As we recalled above, this is a global symmetry. A global symmetry
relating two points $u$ and $u'$ implies that the two corresponding quantum theories
are equivalent and must have the same physical content. In particular, the mass
spectrum with its degeneracies 
must be the same at $u$ and $u'$. Since $m=\sqrt{2}\, \vert an_e-a_Dn_m\vert$,
this implies that for each BPS state $p=(n_e,n_m)$ at $u$ there exists a BPS state
$p'=(n_e',n_m')$ at $u'$ such that
$$\vert\eta(\Omega(u'),p')\vert \equiv \vert a(u')n_e'-a_D(u')n_m'\vert
=\vert a(u)n_e-a_D(u)n_m\vert \equiv \vert\eta(\Omega(u),p)\vert  \ .
\eqn\qia$$
This implies that there exists a matrix $G\in Sp(2,{\Z})$ and a phase
$e^{i\omega}$ such that
$$\pmatrix{a_D\cr a\cr}(u')=e^{i\omega}\, G
\pmatrix{a_D\cr a\cr}(u) \ , \,\pmatrix{n_e'\cr n_m'\cr}=G\pmatrix{
n_e\cr n_m\cr} \ ,
\eqn\ti$$ 
since then  $\vert\eta\bigl(\Omega(u'),p'\bigr)\vert=
\vert\eta\bigl(G\Omega(u),Gp\bigr)\vert = \vert\eta\bigl(\Omega(u),p\bigr)\vert$. 
Thus, if the BPS state
$(n_e,n_m)$ exists at $u$, the BPS state\foot{ 
Of course, by $G (n_e,n_m)$ we mean $G\pmatrix{ n_e\cr n_m\cr}$, but it is
typographically more convenient to write $G (n_e,n_m)$. We will adopt this convention in
the following.
}
$(n'_e,n'_m)=\pm G (n_e,n_m)$ must also exist
at $u'$ with the same mass. The sign has no
importance since $-(n_e,n_m)$ is the antiparticle of \pp\ and is
always present with \pp .

The phase in \ti\ may be surprising. It shows that the relation
between $\Omega (u)$ and $\Omega (u')$ is not in general a duality
transformation. Nevertheless, this sort of new $U(1)$ clearly is a
symmetry of the lagrangian $\cal L$ and of the metric $ds^2$  
which are invariant under the change
$A\to e^{i\omega}A$ and $A_D\to e^{i\omega}A_D$. This amounts to 
performing the transformation ${\cal F}(a)\to e^{2i\omega}{\cal F}
(e^{-i\omega}a)$.
We will see that such a phase does indeed arise
for the \ZZ-symmetry. Before doing so, however, we need to clarify the mathematical
description of the BPS states.

\section{The mathematical description of BPS states}

\fig{Taking $u$ to $u'=-u$ inside the strong-coupling region ${\cal R}_S$ 
one has to cross the cut on $[-1,1]$.}{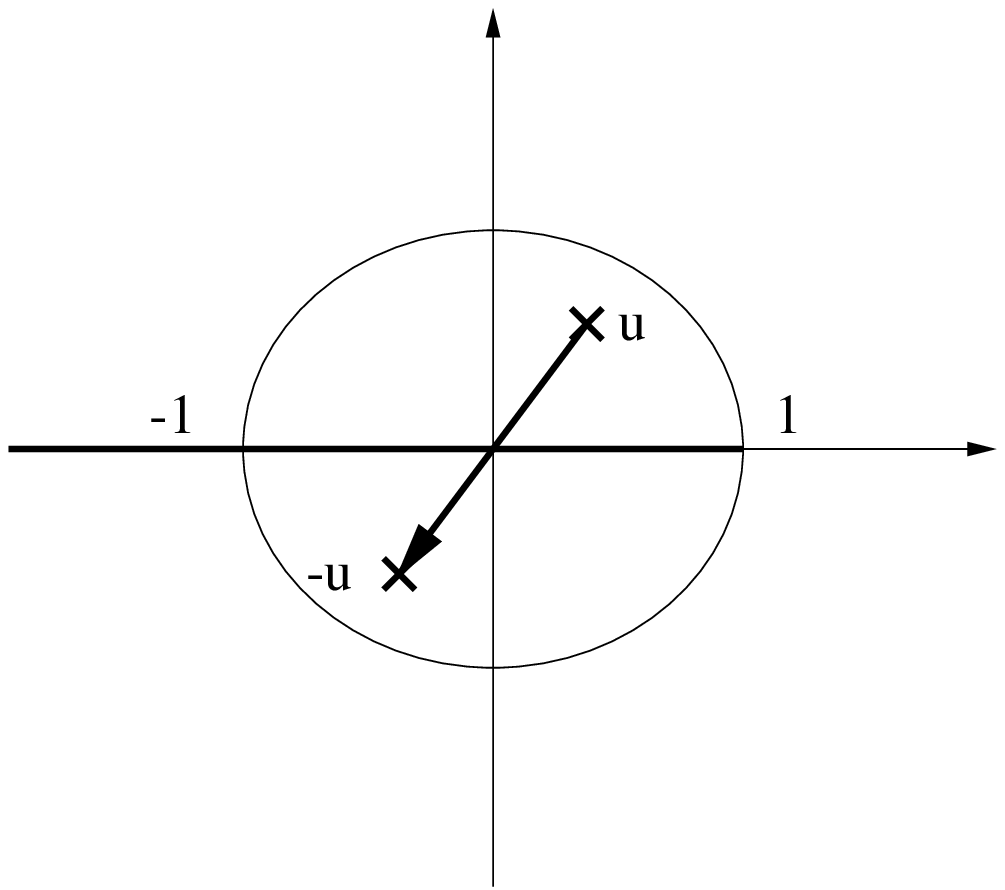}{8cm}
\figlabel\figiv
\vskip 2.mm
As we have discussed in the previous section, the curve \C\ separates a weak-coupling
region ${\cal R}_W$ (outside \C) from a strong-coupling region ${\cal R}_S$ (inside \C).
We already mentioned that the physical spectrum (by which we mean the set of BPS
states, not the mass spectrum)
does not depend on the point $u$ inside a given region ${\cal R}_W$ or ${\cal R}_S$. 
This means that
if a locally constant section $p$ representing a BPS state exists
at $u\in {\cal R}_S$ ($u\in {\cal R}_W$), it will exist at any other point 
$u'\in {\cal R}_S$ ($u'\in {\cal R}_W$).
However, in the strong-coupling region ${\cal R}_S$, 
the section cannot be represented by a {\it unique} couple of
integer numbers $(n_e,n_m)$ through all  ${\cal R}_S$.
We have already encountered the example of the dyon which becomes massless at $u=-1$
and which is represented as $(1,-1)$ or $(1,1)$ depending on whether one approaches
$u=-1$ from the upper or lower half-plane.
This is a consequence of the presence of the singularities and branch cuts (see Fig.
\figi) which
prevent the bundle $E$ from being trivial. To see this, pick
a section $p$ represented by \pp\ at $u\in {\cal R}_S \cap {\H}_+
={\cal R}_{S,+}$ 
where ${\H}_+$ is the upper half-plane 
(${\cal R}_{S,-}$ is defined similarly). The mass of the
BPS state associated with $p$ will be 
$m_p(u)=\sqrt{2} \vert a(u)n_e-a_D(u)n_m\vert$. Now
transport this section through the cut $(-1,1)$ to a point $u'$ in
${\cal R}_{S,-}$ (see Fig. \figiv\ where the case $u'=-u$ is depicted). Of
course the mass $m_p(u)$ will vary continuously in this process, as
physically nothing happens on the cut. 
But once one passes through the cut, 
$m_p$ will no longer be expressed in terms of $a_D$ and $a$ but in
terms of their analytic continuations: 
$m_p(u')=\sqrt{2} \vert \tilde a(u')n_e-\tilde a_D(u')n_m\vert$, where
$(\tilde a_D,\tilde a)(u')=M_1 
(a_D, a) (u')$. One has then $m_p(u')=
\sqrt{2} \vert a(u')\tilde n_e-a_D(u') \tilde n_m\vert$ 
with $(\tilde n_e, \tilde n_m)=
M_1^{-1} (n_e,n_m)$. Hence, the section $p$ will
be represented in ${\cal R}_{S,-}$ by $(\tilde n_e,\tilde n_m)=
M_1^{-1} (n_e,n_m)=(n_e,2n_e +n_m)$. This transformation
insures the continuity of the mass of the state.  Note that the different descriptions
of the same state in terms of different couples of integers is consistent with the
notion of stability. Indeed, if $n_e$ and $n_m$ are relatively prime, then it follows
from B\'ezout's theorem
that any $n_e'$ and $n_m'$, obtained through an
$Sp(2,\Z)$ transformation 
from $n_e$ and $n_m$, are also relatively prime.

We have learned that, though there is a unique spectrum
${\cal S}_S$ valid through all the region ${\cal R}_S$, we must
introduce two different sets of couples \pp\ to represent it.
We will denote these two sets by ${\cal S}_{S,+}$ and ${\cal S}_{S,-}$.
We have:
$$
{\cal S}_{S,-}=M_1^{-1}({\cal S}_{S,+})\ , \quad {\cal S}_{S,+}=M_1({\cal S}_{S,-}) \ .
\eqn\ci$$

In the weak-coupling region ${\cal R}_W$ the situation is simpler. Since any two points
$u,\, u' \in {\cal R}_W$ can be joined by a path  not crossing a cut, for such a path
$a_D$ and $a$ at $u'$ are always given by the same branch as the one at $u$ (the
principal branch). Hence, a section $p$ can be represented by the same couple of
integers through all of ${\cal R}_W$. However, if one wants to compare two sections just
below and above the cut $(-\infty,-1]$, one again needs  to compare different
representations, this time related by $M_\infty$. In particular, for the dyon which
becomes massless at $u=-1$ one has $(1,1)=M_\infty^{-1} (1,-1)$, in analogy with the
first relation \ci.

\section{The \ZZ-symmetry}

\fig{Taking $u$ to $-u$ in the weak-coupling region ${\cal R}_W$ without crossing the
cuts on $(-\infty,1]$}{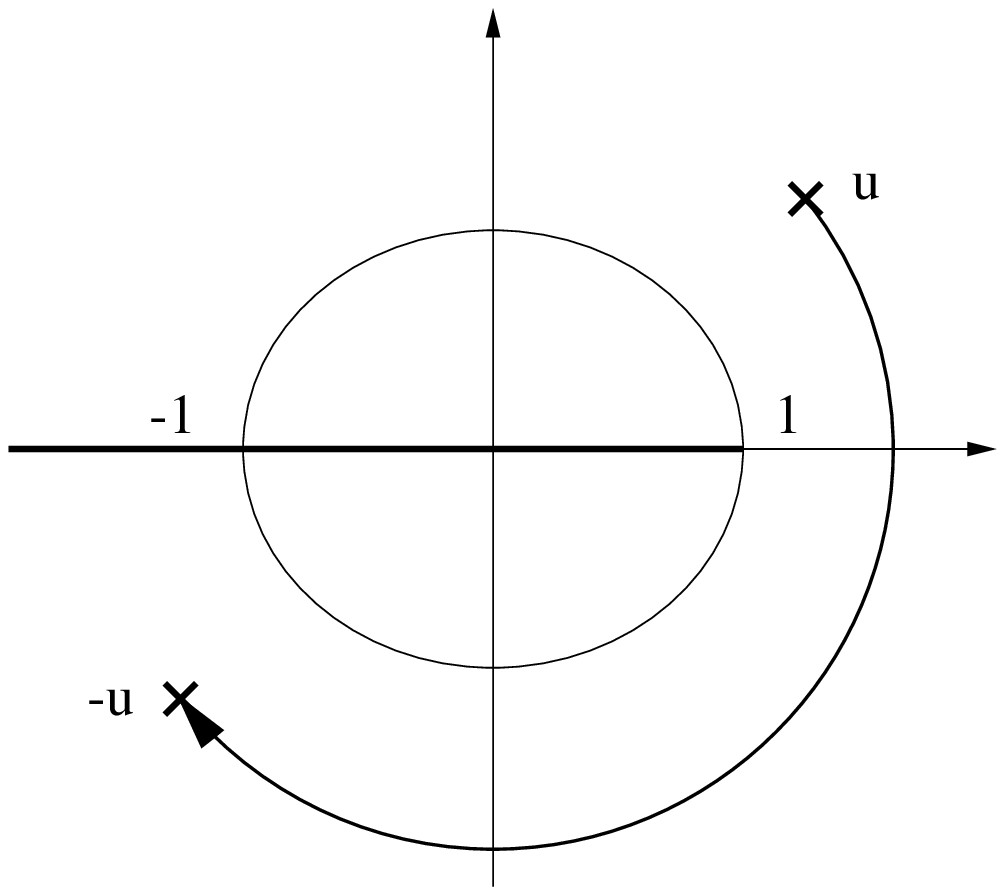}{8cm}
\figlabel\figiii
\vskip 2.mm
Consider now the \ZZ-symmetry $u\to u'=-u$. To start with, we take $u$ in the
upper half-plane and outside the curve \C, {\it i.e.} in the weak-coupling region
${\cal R}_W$. Then $a(-u)$ and $a_D(-u)$ are obtained by analytical continuation along
the path in ${\cal R}_W$ shown in Fig. \figiii\ which does not cross any of the cuts on $(-\infty,1]$.
Using Kummer's relations [\ERD] between hypergeometric functions, namely
$$(1-z)^{c-b-1}\, F(1-a,b+1-c,2-c\, ;{z\over z-1})= F(a+1-c,b+1-c,2-c\, ;z)
\eqn\kummeri$$
with $a={1\over 2}$, $b=-{1\over 2}$, $c=1$, and
$$\eqalign{
e^{i\pi(a+1-c)\tilde\e }\ &{\Gamma(a+1-c)\Gamma(b)\over \Gamma(a+b+1-c)}
F(a,b,a+b+1-c\, ;1-z)\cr
&\qquad {}={\Gamma(a+1-c)\Gamma(1-a)\over \Gamma(2-c)}\ z^{1-c}\, (1-z)^{c-a-b}\,
F(1-a,1-b,2-c\, ;z)\cr
&\qquad\,\, {}+e^{i\pi(a+1-c)\tilde\e }\ {\Gamma(b)\Gamma(1-a)\over \Gamma(b+1-a)}\
(1-z)^{-b}\, F(b,c-a,b+1-a\, ;{1\over 1-z})\cr }
\eqn\kummerii$$
with $a=b={1\over 2}$, $c=0$, and $\tilde\e={\rm sgn}(\IM  z)$,
it is easy to show that
$$\pmatrix{a_D\cr a\cr}(-u)=-i\epsilon \ 
G_{W,\epsilon}\, \pmatrix{a_D\cr a\cr}(u)\ , \quad \quad
G_{W,\epsilon} = \pmatrix{1 &\epsilon\cr 0 &1\cr},
\eqn\tii $$
where $\epsilon$ is the sign of $\IM  u$. Note that
$G_{W,+}=G_{W,-}^{-1}$, as required by consistency.

When determining the corresponding matrices in the strong-coupling region ${\cal R}_S$
one has to be careful about picking the correct analytic continuation; 
since then $u$ is inside the curve \C\ (see Fig. \figiv), when going from $u$
to $-u$ one either has to go through a cut or cross \C\ twice. Since the BPS spectrum
changes when crossing the curve \C\ we have to  go through the cut  instead
(see Section 5.2). We will then use the analytic continuation
of $a_D$ and $a$ through the cut $\lbrack -1,1\rbrack $, that is
$M_1
(-iG_{W,+})\Omega(u)$ if $\IM\, u>0$ and $(M_1)^{-1} (+iG_{W,-}) \Omega(u)$ if
$\IM\, u<0$, so that in the strong-coupling region ${\cal R}_S$ one has
$$G_{S,\epsilon}=(M_1)^{\epsilon}\ G_{W,\epsilon}=
\pmatrix{1 &\epsilon\cr -2\epsilon &-1\cr}\ .
\eqn\tiii$$
Note that 
$$(G_{S,\epsilon})^2=-{\bf 1} \ ,
\eqn\tiv$$
as well as $M_1^{-1}G_{S,+} M_1=-G_{S,-}$, so that $G_{S,-}=-G_{W,+} M_1$ and
$G_{S,+}=-G_{W,-} M_1^{-1}$.

One could ask if some new $G$ matrices occured when going from $u$
to $-u$ in the region ${\cal R}_W$ crossing the cut 
$( -\infty ,-1\rbrack$. The answer is no, as the matrices
we obtain in this case are $G_{W,\epsilon}' =
(M_\infty)^\e\, G_{W,\e} =-G_{W,-\epsilon}$.
Finally, let us mention that the democracy transformation
amounts to conjugating the $G$ matrices by $M_{\infty}^p$, exactly
as for the monodromy matrices.
\chapter{The spectrum of BPS states}

\section{The weak-coupling spectrum}

Here we will prove that
${\cal S}_W$ is composed of\foot{
To be precise, there is also the everywhere massless abelian
$N=2$ vector multiplet with
$n_e=n_m=0$, describing the photon, etc., which is the only one appearing in the
low-energy effective action. This multiplet is present on all of ${\cal M}$, but it has
a status very different from all other BPS states since it is a vector and not a
hyper-multiplet.
}
the massive gauge bosons $\pm (1,0)$,
usually called ${\rm W}^\pm$, and the
dyons $\pm (n,1),\, n\in\Z$. We insist on the fact that
our method of proof is completely different from the usual
semiclassical approach. It relies on the knowledge of the 
low energy wilsonian action only, {\it i.e.} the solution \div\ with its \ZZ-symmetry, 
and involves very simple arguments.

The states $\pm (1,0)$ are in the perturbative spectrum of the theory
and belong trivially to ${\cal S}_W$. Note that they are invariant under the
\ZZ-symmetry since $G_{W,\e} (1,0)=(1,0)$. Moreover, as seen above,
the monopole and antimonopole $\pm (0,1)$ also are in ${\cal S}_W$. At a point
$u\in {\cal R}_W$, this monopole has a mass 
$m(u)=\sqrt{2}\,\vert a_D(u)\vert =
 \sqrt{2}\,\vert \eta \bigl( (0,1),  \Omega(u)\bigr)\vert$.
As discussed above, the
same section must also exist at $-u$, since it can be transported along the path of Fig.
\figiii, but with a mass $m(-u)=
 \sqrt{2}\,\vert \eta \bigl( (0,1),  \Omega(-u)\bigr)\vert
=\sqrt{2}\,\vert \eta \bigl( (0,1), G_{W,\e} \Omega(u)\bigr)\vert
=\sqrt{2}\,\vert \eta \bigl(  G_{W,\e}^{-1} (0,1),\Omega(u)\bigr)\vert$. 
By the \ZZ-symmetry, at $u$, there must exist a state
$G_{W,\epsilon}^{-1} (0,1)=(-\epsilon,1)$ which has the same mass 
(at $u$) as $(0,1)$ has at $-u$. (Recall that $\e=\pm$ is always the sign of
$\IM u$). This proves the existence of the dyon $(-\e,1)$ at $u$. 
Repeating this reasoning, the \ZZ-symmetry and the existence of the dyon $(-\e,1)$ at
$u$ implies the existence of a state $G_{W,\epsilon}^{-1} (-\e,1)=(-2\epsilon,1)$ at
$u$, and hence everywhere in ${\cal R}_W$. Since $\e=\pm 1$, depending on where one
started in ${\cal R}_W$, it follows by induction that all dyons $(n,1),\, n\in\Z$ exist
in ${\cal R}_W$. In other words,
${\cal S}_W$ is invariant 
under the transformations generated by the matrices
$G_{W,\epsilon } $: 
$${\cal S}_W = G_{W,\epsilon }({\cal S}_W) \ ,
\eqn\ca$$
and all the dyons $\pm (n,1),\, n\in\Z$ indeed belong
to ${\cal S}_W$.

Next, let us show that there are no other states in ${\cal S}_W$. Suppose
\pp\ is in ${\cal S}_W$. We exclude the case $n_m=0$ since this is either the
W$^\pm$-boson $\pm (1,0)$
which we know is part of ${\cal S}_W$ or, if $n_e\ne \pm 1$, an unstable state.
Then, as before, the \ZZ-symmetry
implies that all the states generated by $G_{W,\pm}$ from $(n_e,n_m)$, {\it i.e.} all
the states of the form $(n_e+k n_m,n_m),\, k\in{\Z}$,
are also in ${\cal S}_W$. Of course, there always exists a $k_0\in\Z$
such that $(n_e+k_0n_m)/n_m=n_e/n_m +k_0\in\lbrack -1,1\rbrack$.
Thus $(n_e+k_0 n_m,n_m)$ will become massless at the point $u^*$ on \C\ where
$(a_D/ a)(u^*)= n_e/n_m +k_0$ and  hence must equal $\pm (0,1)$, $\pm (1,1)$ or 
$\pm (-1,1)$. In all cases
this implies that $n_m=\pm 1$ and thus \pp\ is one of the states $\pm (n,1)$.

The argument is not modified after a democracy transformation. The
$G$ matrices relevant here are not changed since $M_{\infty}^p
G_{W,\epsilon}M_{\infty}^{-p}= G_{W,\epsilon} $. After a democracy transformation
$a_D/a$ will run from $-1-2p$ to $1-2p$ on the curve
$\cal C$, so that the states that become massless are
$\pm (-2p,1)$ and $\pm (1-2p,1)\equiv\pm (-1-2p,1)$. It is clear that exactly the same
spectrum ${\cal S}_W$ is generated from $(-2p,1)$ and $\pm (1,0)$.

\section{The strong-coupling spectrum}

Now take a section $p$ in the strong-coupling spectrum, which at a point $u\in {\cal
R}_{S,+}$ is represented by $(n_e,n_m)\in {\cal S}_{S,+}$. It should now  
be clear that, by the \ZZ-symmetry, the state $G_{S,+} (n_e,n_m)$ then also is in ${\cal
S}_{S,+}$. However, since the argument in the strong-coupling region, with its
distinction between ${\cal S}_{S,+}$ and ${\cal S}_{S,-}$, is potentially more confusing,
we will give the argument in detail again.

Since one can go from $u$ to $-u$ without crossing the curve \C, the same section $p$
must also exist at $-u$, but is represented by $(\tilde n_e, \tilde n_m) = M_1^{-1}
(n_e,n_m)$ according to our discussion in Section 4.2. At $-u$ this state then has a
mass 
$$\eqalign{m_p(-u)=\sqrt{2}\, \vert \tilde n_e a(-u)-\tilde n_m a_D(-u) \vert 
&=\sqrt{2}\, \vert \eta\bigl(M_1^{-1} (n_e,m_m),\Omega(-u)\bigr)\vert\cr 
&=\sqrt{2}\, \vert \eta\bigl((n_e,n_m),M_1\Omega(-u)\bigr)\vert\cr} 
\eqn\cdiii$$
which by eq. \tiii\ equals
$$\sqrt{2}\,\vert \eta\bigl((n_e,n_m),G_{S,+}\Omega(u)\bigr)\vert 
=\sqrt{2}\,\vert \eta\bigl((G_{S,+})^{-1}(n_e,n_m),\Omega(u)\bigr)\vert  \ .
\eqn\cdiv$$
By the \ZZ-symmetry there must be a section $p'$, 
represented by $(n_e',n_m')$ in ${\cal
R}_{S,+}$, 
which at $u$ has the same mass as $p$ has at $-u$, {\it i.e.}
$$p'=G_{S,+}^{-1}\, p = - G_{S,+}\, p \ .
\eqn\cdvi$$
We conclude that if $p\in {\cal S}_{S,+}$ then also $G_{S,+}\, p\in {\cal S}_{S,+}$.
The same applies for 
$ {\cal S}_{S,-}$. Hence
$$ {\cal S}_{S,\pm}=G_{S,\pm}( {\cal S}_{S,\pm}) \ .
\eqn\cii$$

Now we are in a position to determine the strong-coupling spectrum ${\cal S}_S$. We know
that the magnetic monopole, becoming massless on the curve \C\ at $u=1$, must exist in
${\cal R}_S$ and hence be in ${\cal S}_S$. Since $M_1 (0,1)=(0,1)$ it is described by
the same couple of integers in ${\cal R}_{S,+}$ and ${\cal R}_{S,-}$. Let us determine
${\cal S}_{S,+}$ first. Take $p$ in eq. \cdvi\ to be the monopole. Then $p'=G_{S,+}\, p$
is the dyon $(1,-1)$. Applying $G_{S,+}$ again yields $-(0,1)$, and
hence gives back the monopole. This is very different from the weak-coupling spectrum
where all dyons (with unit magnetic charge) are generated from the monopole. Actually,
since $(G_{S,\e})^2=-1$, all BPS states in ${\cal S}_{S,+}$ come in \ZZ-pairs. For a
general stable BPS state described by $(n_e,n_m)\in {\cal S}_{S,+}$ the \ZZ-pairs are
$$(n_e,n_m)\in {\cal S}_{S,+}\ \  \Leftrightarrow\ \  
G_{S,+}(n_e,n_m) =(n_e+n_m, -2n_e-n_m)
\in {\cal S}_{S,+} \ .
\eqn\ciii$$
We will now show that for each of the \ZZ-pairs one or the other member becomes
massless somewhere on \CP. Since we know that $\pm(0,1)$ and $\pm(1,-1)$ are the only
states in the physical spectrum 
that become massless, we then conclude that this is the only pair in 
$ {\cal S}_{S,+}$. 

Recall that \CP\ is the part of \C\ in the upper half-plane, which is
the only part seen from ${\cal R}_{S,+}$, and that ${a_D\over a}(u)$ takes all real
values in $[-1,0]$ on \CP. First consider the case $n_m=0$. Then, for stability, one
has $n_e=1$ (up to an irrelevant sign). By \ciii, the \ZZ-transformed state is
$(1,-2)$. At the point $u\in\ $ \CP\ such that  ${a_D\over a}(u)=-{1\over 2}$, 
this state
becomes massless. Hence $(1,0)$ and $(1,-2)$ cannot be in $ {\cal S}_{S,+}$.
The nonexistence of the W-bosons $(1,0)$ in the strong-coupling region has been
suggested before in [\ROC ] using completely different arguments. 
Next, let
$n_m\ne 0$. Then $(n_e,n_m)$ will become massless on \CP\ if there is a point
$u\in\ $ \CP\ where  ${a_D\over a}(u)={n_e\over n_m}\equiv r$, {\it i.e.} if $r\in
[-1,0]$. The \ZZ-partner will
become massless on \CP\ if there is a point $u'\in\ $ \CP\ where 
${a_D\over a}(u')=-(n_e+n_m)/(2n_e+n_m)=-(r+1)/(2r+1)\equiv \varphi(r)$, {\it
i.e.} if $\varphi(r)\in [-1,0]$. It
is easy to see from the properties of the function $\varphi(r)$
that one or the other case is always realised, {\it i.e.} either 
$r\in [-1,0]$ (and then $\varphi(r)\notin (-1,0)$)
or $\varphi(r)\in (-1,0)$ (for $r\notin [-1,0]$).
So one or the other \ZZ-partner always becomes massless on \CP, and we conclude that
$${\cal S}_{S,+}=\left\{ \pm (0,1)\, ,\ \pm (1,-1)\right\} \ .
\eqn\civ$$
Exactly the same reasoning applies to ${\cal S}_{S,-}$ with $G_{S,-}$ replacing
$G_{S,+}$. But ${\cal S}_{S,-}$ is most easily determined by using
eq.~\ci , and we obtain
$${\cal S}_{S,-}=\left\{ \pm (0,1)\, ,\ \pm (1,1)\right\} \ .
\eqn\cv$$
Recall that $(1,-1)\in{\cal S}_{S,+}$ and $(1,1)\in{\cal S}_{S,-}$ are the two different
descriptions of the same section $p$ corresponding to one and the same dyon. So the
strong-coupling spectrum contains exactly two BPS states.

Let us remark that a democracy transformation does not affect these conclusions, except
that, for ${\cal S}_{S,+}$ for example, 
the sections corresponding to $(0,1)$ and $(-1,1)$ are now
described by $(-2p,1)$ and $(-1-2p,1)$, and that ${a_D\over a}(u) \in
[-1-2p,-2p]$ on \CP, and \penalty -1000 $\varphi(r) \to 
\varphi_p(r)=-\bigl((1+4p)r+8p^2+4p+1\bigr)/\bigl(2r+4p+1\bigr)$ 
\penalty -1000 \ with either $r\in [-1-2p,-2p]$, or if
$r\notin [-1-2p,-2p]$ then $\varphi_p(r)\in (-1-2p,-2p)$.
In any case, the strong-coupling spectrum precisely consists of the two sections
describing the two BPS states that become massless at $u=1$ or $u=-1$.

\chapter{Crossing the curve $\cal C$}

We have seen that the strong-coupling spectrum only contains $\pm(0,1)$ and $\pm(1,-1)$
in ${\cal R}_{S,+}$ and only $\pm(0,1)$ and $\pm(1,1)$
in ${\cal R}_{S,-}$. A nice physical picture of this is that 
any BPS state in the weak-coupling spectrum,
$\pm(1,0)$ or $\pm (n,1)$, has to decay into these two states when crossing the curve
\C. We will now illustrate how this goes.\foot{
Of course, the kinematic possibility of the decay reactions mentioned in this section
is well-known, see {\it e.g.} [\FAY]. 
}

Suppose one crosses the curve \C\ at the point $u^*$ in the upper half-plane, {\it i.e.}
on \CP, where ${a_D\over a}(u^*)=r\in [-1,0]$. Start with a dyon $(n,1)$ with $n>0$. By
conservation of the electric and magnetic charges the decay reaction must be
$$(n,1)\ \to \ n\times (1,-1) + (n+1)\times (0,1) \ .
\eqn\si$$
The masses of $(n,1)$, $(1,-1)$ and $(0,1)$ at $u^*$ are
$\sqrt{2}\, \vert n a(u^*)-a_D(u^*)\vert
=\sqrt{2}\, \vert  a(u^*)\vert\ \vert n-r\vert$, 
$\sqrt{2}\, \vert  a(u^*)\vert\ \vert 1+r\vert$ and
$\sqrt{2}\, \vert  a(u^*)\vert\ \vert r\vert$, and the decay is possible (and does take
place) since one has the conservation of total mass:
$$ \vert n-r\vert = n+\vert r\vert = n\times (1-\vert r\vert ) + (n+1)\times \vert r
\vert = n\times \vert 1+r\vert + (n+1)\times \vert r\vert  \ .
\eqn\sii$$
For $n<0$, the decay reaction is
$(n,1)\ \to \ \vert n\vert \times (-1,1) + (\vert n\vert -1)\times (0,-1) \ ,
$
(where $(-1,1)$ and $(0,-1)$ are the anti-dyon and anti-monopole)
with the masses working out similarly. The W-bosons $(\pm 1,0)$ decay as
$(\pm 1,0)\ \to \ (\pm 1, \mp 1) + (0,\pm1)   \ ,
$
with the mass balance given by
$1=\vert 1+r\vert + \vert r\vert 
$
which is satisfied since $-1\le r\le 0$.

When one crosses \CM\ instead of \CP, $r\in [0,1]$ is positive instead, and $(\pm 1,\mp
1)$ is replaced by $\pm (1,1)$, so that everything works out exactly the same way. Also
the decay of anti-dyons $-(n,1)$ is exactly the mirror of the decay of the dyons $(n,1)$.

An alternative way of studying the strong-coupling spectrum may be to compute
all the possible decays of the states belonging to the weak coupling spectrum
into arbitrary states $(n_e,n_m)$,
at any point on $\cal C$. Only those states which can be produced by such
a process at {\it all} the points on $\cal C$ can eventually be present in
${\cal S}_S$. This seems to be a strong constraint, and the monopole and the dyon
may well be the unique states having this property. 
However, a proof of this fact doest not seem 
to exist yet. Moreover, it is impossible with this method to 
prove that a state {\it a priori} of marginal stability actually does decay. 
The main ingredient we used to overcome this difficulty is the global 
\ZZ-quantum symmetry on the moduli space.

\chapter{Conclusions and outlook}

Let us recapitulate our assumptions:
\pointbegin
$a(u)$ and $a_D(u)$ are given by the Seiberg-Witten solution \div.
\point
A charged massless BPS state at a 
point $u\in {\cal M}$ leads to a singularity in the low-energy
effective action and hence in $a(u)$ or $a_D(u)$, and thus there are no other 
charged massless
states than those associated with the two singularities at $u=\pm 1$.
\point
The \ZZ-symmetry is a true quantum symmetry acting on the moduli space as $u\to -u$.
\point
The mass of a BPS state is given by $m=\sqrt{2}\,\vert n_e a-n_m a_D\vert$.
\par
\noindent
Of course, as physicists we believe that all of these assumptions are true. In any case,
assuming them to be valid, we did show that
\pointbegin
the weak-coupling spectrum is the well-known one composed  of the dyons $\pm (n,1)$ and
the W-bosons $\pm (1,0)$, and
\point
the strong-coupling spectrum contains only the two BPS  particles that can become
massless and are responsible for the singularities of the Seiberg-Witten solution: the
monopole $(0,1)$ and the dyon, described either as $(1,1)$ or as $(1,-1)$ (as well as
their antiparticles, of course).

One may speculate on what happens for the generalisations to 
gauge groups other than
$SU(2)$ and to theories including extra matter. 
We are tempted to conjecture that there, too, the strong-coupling spectra
consist of those BPS states that become massless at the singularities in moduli
space. However, the structure of the moduli space is much more complex than the one
studied here, and a detailed investigation clearly is necessary.

\noindent{\bf Note Added}

By now, we have confirmed this conjecture for the $N=2$ $SU(2)$ theories with one, two
or three massless quark hypermultiplets [\BF].

\refout
\end